# Influence of elastic deformations on body-wave velocity in solids: a case study considering shear deformations in concrete

Hao Cheng,[1,*] Cornelis Weemstra,[2,3] Katrin Löer,[3] Max A.N. Hendriks,[1,4] and Yuguang Yang[1]

[1] Department of Engineering Structures, Faculty of Civil Engineering and Geosciences, Delft University of Technology, 2628 CN Delft, the Netherlands
[2] Royal Netherlands Meteorological Institute (KNMI), 3731 GA De Bilt, the Netherlands
[3] Department of Geoscience and Engineering, Faculty of Civil Engineering and Geosciences, Delft University of Technology, 2628 CN Delft, the Netherlands
[4] Department of Structural Engineering, Faculty of Engineering, Norwegian University of Science and Technology, 7491 Trondheim, Norway
[*] Email: h.cheng-2@tudelft.nl

**Abstract**
This paper investigates the influence of elastic deformation on the velocity of body waves in compressible isotropic materials making use of the framework of acoustoelasticity. Specifically, it examines body waves propagating at an angle to the principal deformation axes, where both shear and normal deformations are present in the coordinate system defined by the wave propagation direction. While numerous efforts have addressed this topic, the theoretical derivations have not yet to provide definitive conclusions about the response of wave velocity to applied shear stresses and strains. To derive more specific conclusions for body waves in concrete, we analyzed three examples using concrete as the medium. The key findings are that, in case of concrete materials when body waves propagate on the shear deformation plane, variations in longitudinal wave velocity are predominantly attributed to changes in normal strains, whereas transverse wave velocity is significantly influenced by both normal and shear strains. This finding can enhance the use of acoustoelasticity for detecting the magnitudes and directions of principal stresses in plane stress state applications.

## 1. Introduction

The acoustoelastic effect, named by Toupin and Bernstein [1], refers to the change in velocity of an elastic wave within a (pre-)deformed elastic medium. The earliest investigations on this topic can be traced back to the 1820s [2]. In the 1950s and 1960s, researchers established the fundamental framework of acoustoelastic theory by incorporating the nonlinear terms into the constitutive equations of elastic materials [1, 3-5]. The framework is validated through experimental studies using both body waves and surface waves [3, 6].

However, the current expressions derived from the acoustoelastic framework is limited to cases where the wave propagation direction is either parallel or perpendicular to one of the principal axes of deformation. In practical terms, if the wave does not propagate along these directions, these expressions are not applicable. This constraint limits the widening of application scenarios for acoustoelasticity in engineering practice outside laboratory, in which, researchers can design test setups allowing the alignment of the sensor layout with the wave propagation direction. With the recent development in sensor technology, long-term monitoring utilizing permanently deployed ultrasonic sensors within large concrete infrastructure has become a practical option. However, in these applications, there is no assurance that elastic waves will consistently propagate along these predetermined directions. Consider the scenario of monitoring a concrete bridge deck under the influence of moving vehicle loads, where the stress condition changes over time. In such an instance, it becomes impractical to consistently align the wave propagation direction along the principal deformation directions.

When there is an angle between the body wave propagation direction and the principal stress direction, by establishing a new coordinate system aligned parallel and perpendicular to the wave propagation direction and computing the stress matrix within this new coordinate system, shear stresses can be identified. Simultaneously, the presence of shear stresses concurrently indicates the existence of shear deformations, which is defined as an isochoric plane deformation in which there are a set of line elements with a given reference orientation that do not change length and orientation during the deformation [7]. In this case, incorporating shear stresses/strains into the framework of acoustoelasticity is crucial for expanding its range of applicability.

Encompassing shear stresses/strains within the realm of acoustoelasticity is not a new concept. Extensive research has been conducted by Boulanger & Hayes [8] and Ogden and his colleagues [7, 9-14], whose theoretical framework can be extended to incorporate shear stresses and strains. The theoretical work presented by these authors is of high quality and demonstrates meticulous rigor. Nevertheless, the complexity of their expressions may pose difficulties for engineers when applying them to practical engineering scenarios, and no specific conclusion regarding the response of wave velocity to applied shear stresses and strains can be drawn from the theoretical derivations.

This paper aims to investigate the influence of elastic deformations on body-wave velocity in compressible isotropic solids and further simplifies its application in detecting stress and strain changes. To achieve this, a general acoustoelastic theory that incorporates both normal and shear deformations is introduced in Section 3. Three numerical examples using concrete as the medium are presented in Section 4 to explore the influence of normal and shear strains on body-wave velocity. Section 5 discusses how the findings in this paper could be applied to structural health monitoring in the future, as well as the impact of concrete heterogeneity on acoustoelasticity.

## 2. Brief introduction on the development of acoustoelastic theory
### 2.1 Linear elasticity-based acoustoelastic theory
The theory of acoustoelasticity elucidates the relationship between applied stress and elastic wave velocity within the framework of continuum mechanics. It considers infinitesimal dynamic disturbances superimposed on a large, elastically deformed body [15], a concept known as *small-on-large*. Originally proposed by Cauchy [2] in 1829, this concept was further developed by Rayleigh [16], Brillouin [17], and Biot [18]. During this development, the connection between wave velocity and stress is established in the realm of linear elasticity. However, compared to the much smaller strain due to wave motion, the initial static strain is significantly larger and cannot be considered to be infinitesimal in the constitutive equation. Therefore, the assumption of linear elasticity becomes inadequate. A natural progression from this stage involves proposing a stress-strain relationship that accounts for higher orders of deformation.
### 2.2 Development of nonlinear elasticity
The first work addressing non-linear elasticity is attributed to Brillouin [17]. his main conclusion still falls within the realm of linear elasticity, since *on peut montrer que ces termes ne modifient* *pas la vitesse de propagation tant qu'il s'agit d'ondes de très petites amplitudes* (it can be shown that these terms [third order elastic constants] do not modify the propagation speed as long as the waves have very small amplitudes).

The most widely applied form of nonlinear elasticity in the current acoustoelasticity literature originates from Murnaghan's derivation [19]. Assuming energy conservation, Murnaghan demonstrates the necessity of six constants to describe strain energy in compressible, isotropic materials: one representing initial hydrostatic pressure, two second-order elastic constants (Lamé constants), and three third-order elastic constants denoted as *l*, *m*, and *n*, commonly known as *Murnaghan constants*. Subsequently, many researchers contributed to the field of nonlinear elasticity by proposing general forms of the strain-energy function without identifying third-order elastic constants. Notable contributors include Rivlin [20, 21], who focused primarily on incompressible isotropic materials [22-27], and Green and Shield [28], who also focused on these materials.

Please note that Murnaghan constants are not the only third-order elastic constants documented in literature. For instance, Landau and Lifshitz's '*Theory of Elasticity*' acknowledges the existence of third-order elastic constants, though they were presented as an exercise for readers to derive [29]. Toupin and Bernstein [1] introduce $\upsilon_1$, $\upsilon_2$ and $\upsilon_3$ as third-order elastic constants, referring to them as *third-order Lamé constants*. Furthermore, Johnson [5] discusses a different set of third-order elastic constants, denoted as $\beta_1$, $\beta_2$ and $\beta_3$.

### 2.3 Introduction of non-linear elasticity to acoustoelastic theory

One of the earliest attempt to incorporate nonlinear elasticity into the acoustoelastic theory was made by Green et al. [30]. They proposed a general theory for small elastic deformations of an isotropic elastic body superposed on a known finite deformation. Since the strain-energy function is not specified, the theory is applicable to both compressible and incompressible materials. Hayes and Rivlin [31] further discussed the restriction on the strain-energy function to ensure its validity for a material capable of sustaining a state of pure, homogeneous deformation. However, they did not propose any explicit expressions for acoustoelasticity in their work.

In 1953, Hughes and Kelly [3] incorporated Murnaghan constants into the constitutive equation and derived the explicit expressions for acoustoelasticity in compressible isotropic materials. Similar expressions were subsequently derived by other researchers, including Toupin and Bernstein [1], Thurston and Brugger [4], Johnson [5], and Pao and Gamer [15]. In this paper, the theoretical background follows the notation adopted by Pao and Gamer.

Due to space limitations, numerous theoretical works [32-41] are not mentioned in this section. We extend our sincere respect to these pioneers whose exploration and dedication have propelled the theory and application of acoustoelasticity to its current state.

## 3. Theoretical background

This section will briefly introduce the theoretical background of acoustoelasticity. Given that the primary applications of acoustoelasticity lie in non-destructive testing [6], which often involves compressible materials such as polymers, metals, and concrete, the derivation presented in this paper specifically focuses on the compressible isotropic medium.

## 3.1 Equation of motion

The equation of motion for acoustoelasticity is given as [15]:

$$B_{ijkl} \frac{\partial^2 u_k^{\text{incremental}}}{\partial x_j \partial x_l} = \rho^0 \frac{\partial^2 u_i^{\text{incremental}}}{\partial t^2} \ , \quad (1)$$

where $\rho^0$ are the mass density in the natural state without load applied, and $u_i^{\text{incremental}}$ denotes the displacements of dynamic disturbance. The acoustoelastic modulus tensor $B_{ijkl}$ is:

$$\begin{aligned} B_{ijkl} := & C_{jlmn} e_{mn} \delta_{ik} + C_{ijkl} + C_{ijklmn} e_{mn} \\ & + C_{mjkl} \frac{\partial u_i^{\text{initial}}}{\partial a_m} + C_{imkl} \frac{\partial u_j^{\text{initial}}}{\partial a_m} + C_{ijml} \frac{\partial u_k^{\text{initial}}}{\partial a_m} + C_{ijkm} \frac{\partial u_l^{\text{initial}}}{\partial a_m} \ . \end{aligned} \quad (2)$$

where $e_{mn}$ indicates the static strain tensor caused by the external load, and $u_i^{\text{initial}}$ denotes the displacements associated with the static deformation induced by the external load. The $C_{ijkl}$ in Eq. (2) represents the second-order elastic coefficients composed of the second-order elastic constants, also known as Lamé parameters, $\lambda$ and $\mu$. The second-order elastic coefficients can be expressed using Voigt notation as $C_{IJ}$. The third-order elastic coefficients, $C_{\alpha\beta\gamma\delta\varepsilon\eta}$, can be expressed using the third-order elastic constants, i.e., Murnaghan constants $l$, $m$, and $n$. Similar to $C_{ijkl}$, the third-order elastic coefficients $C_{\alpha\beta\gamma\delta\varepsilon\eta}$ can also be represented using Voigt notation as $C_{KLM}$, which are commonly known as standard third-order elastic coefficients.

## 3.2 Plane wave propagation

To investigate the wave velocities among different wave modes, a plane harmonic wave with the following form is introduced into the equation of motion [3, 15, 42, 43]:

$$u_\gamma = U_\gamma e^{i(k N_\lambda x_\lambda - \omega t)} \ , \quad (3)$$

where $N_\lambda$ is a unit vector normal to the plane wave that relates to the wave propagation direction, $k$ represents the wavenumber, $\omega$ is the angular frequency, and $U_\gamma$ is the amplitude vector indicates the wave polarization direction. The wave mode is dictated by the relation between $N_\lambda$ and $U_\gamma$. When they are orthogonal, indicating that the polarization direction is perpendicular to the wave propagation direction, these waves are identified as transverse waves or shear waves. Conversely, when the vectors are parallel, signifying that the polarization direction aligns with the wave propagation direction, these waves are termed longitudinal waves or compression waves. We will only discuss these two wave modes in the remaining part of this paper. The choice of a harmonic wave is made to simplify the derivation process, but the outcome is also applicable to plane waves with a general time function [15]. Substituting Eq. (3) into Eq. (1) gives:

$$\left[ B_{ijkl} \left( \delta_{j\lambda} N_\lambda \right) \left( \delta_{l\lambda} N_\lambda \right) - \rho^0 v^2 \delta_{ik} \right] U_k = 0 \ , \quad (4)$$

where $v$, coinciding with $\omega/k$, denotes the wave velocity.

## 3.3 Governing equation

To solve Eq. (4) analytically, it is necessary to specify either the propagation vector [5, 15, 44] $N_\lambda$ or the amplitude vector [42] $U_k$ as known. From a structural monitoring (experimental) perspective, it is more practical to specify the known propagation vector. This vector, which points in the propagation direction, is easier to determine in a measurement than the wave polarization-related amplitude vector. Therefore, we adopt this configuration and propose a governing equation based on Eq. (4) with a specified vector $N_\lambda$, **N**=(1;0;0), while leaving the amplitude vector $U_k$ unspecified. Substituting the vector **N** into Eq. (4) gives:

$$\left( B_{i1k1} - \rho^0 v^2 \delta_{ik} \right) U_k = 0 \ . \quad (5)$$

The determination of the wave mode relies on the relation between the propagation vector $N_\lambda$ and the amplitude vector $U_k$. Equation (5) can be expressed in the matrix form as:

$$\begin{bmatrix} B_{1111} & B_{2111} & B_{3111} \\ B_{1121} & B_{2121} & B_{3121} \\ B_{1131} & B_{2131} & B_{3131} \end{bmatrix} \begin{bmatrix} U_1 \\ U_2 \\ U_3 \end{bmatrix} = \rho^0 v^2 \begin{bmatrix} U_1 \\ U_2 \\ U_3 \end{bmatrix}, \tag{6}$$

where the acoustoelastic moduli can be obtained using Eq. (2) and are:

$$\begin{aligned} B_{1111} &= C_{11} + (5C_{11} + C_{111})e_{11} + (C_{12} + C_{112})e_{22} + (C_{13} + C_{113})e_{33} \\ &= \lambda + 2\mu + (5\lambda + 10\mu + 2l + 4m)e_{11} + (\lambda + 2l)(e_{22} + e_{33}) , \end{aligned} \tag{7a}$$

$$\begin{aligned} B_{2121} &= C_{66} + (C_{11} + 2C_{66} + C_{661})e_{11} + (C_{12} + 2C_{66} + C_{662})e_{22} + (C_{13} + C_{663})e_{33} \\ &= \mu + (\lambda + 4\mu + m)e_{11} + (\lambda + 2\mu + m)e_{22} + \left(\lambda + m - \frac{n}{2}\right)e_{33} , \end{aligned} \tag{7b}$$

$$\begin{aligned} B_{3131} &= C_{55} + (C_{11} + 2C_{55} + C_{551})e_{11} + (C_{13} + 2C_{55} + C_{553})e_{33} + (C_{12} + C_{552})e_{22} \\ &= \mu + (\lambda + 4\mu + m)e_{11} + (\lambda + 2\mu + m)e_{33} + \left(\lambda + m - \frac{n}{2}\right)e_{22} , \end{aligned} \tag{7c}$$

$$B_{1121} = B_{2111} = 2(C_{11} + C_{166})e_{12} = (2\lambda + 4\mu + 2m)e_{12} , \tag{7d}$$

$$B_{1131} = B_{3111} = 2(C_{11} + C_{155})e_{13} = (2\lambda + 4\mu + 2m)e_{13} , \tag{7e}$$

$$B_{2131} = B_{3121} = 2(C_{44} + C_{654})e_{23} = \left(2\mu + \frac{n}{2}\right)e_{23} , \tag{7f}$$

where $e_{11}$, $e_{22}$ and $e_{33}$ represent normal strains along $x$-, $y$- and $z$-axes, respectively. Shear strains in the $x$-$y$, $x$-$z$ and $y$-$z$ plane are denoted by $e_{12}$, $e_{13}$ and $e_{23}$, respectively. The derivation details of acoustoelastic moduli $B_{ijkl}$ can be found in Appendix. The second- and third-order elastic coefficients are expressed using the Voigt notations. Please note that in the following sections, we differentiate between strain tensor and principal strains using the following notation: the strain tensor, including normal and shear strains, is represented using $e_{ij}$, while the principal strain is denoted as $e_i$.

Equation (7d) can also be derived using Eq. (12) introduced in the article by Bobrenko et al. [45]. Additionally, normal and shear strains in Eq. (7) can be obtained from principal strains by applying a coordinate transformation. This transformation is based on fixing the propagation direction along the $x$-axis in the new coordinate system and transforming the strain matrix accordingly. We noticed a more elegant approach proposed by Ogden [7], which incorporates the propagation vector into the plane wave form. This approach results in the acoustoelastic moduli tensor, or instantaneous elastic moduli $A_{0ijkl}$, being related to the principal strains, assuming that elastic constants remain unchanged during the loading process. This notation is widely used in later research on acoustoelasticity [46-49]. Shams et al. [50] compared the results obtained using Ogden's notation and with those of Hughes and Kelly [3], finding them to be in agreement.

Equation (6) can be viewed as an eigenvalue equation for the acoustoelastic modulus matrix. The eigenvalues of this matrix are equal to $\rho^0 v^2$, with the corresponding eigenvectors being the amplitude vector $U_k$. Determining the wave velocity necessitates both the eigenvalue and eigenvector: the eigenvector identifies the wave mode, while the eigenvalue provides the velocity of this mode. Thus, eigenvalues and eigenvectors hold equal significance. In the acoustoelastic modulus matrix, the diagonal elements pertain to terms associated with normal strains, while shear strains are hidden within the off-diagonal elements, as depicted in Eq. (7). In the remainder of this paper, the acoustoelastic modulus matrix before and after the coordinate rotation will be referred to as the $B^0$-matrix and $B$-matrix, respectively. This approach, which involves determining the velocity and polarization direction by solving the eigenvalue problem, has also been highlighted by Shams et al. [50] and Pau & Vestroni [51].

## 4. Numerical examples

Based on this governing equation, the influence of body waves velocity due to both normal and shear strains in a medium will be investigated through three practical examples. Example 1 in Section 4. A demonstrates the derivation of analytical solutions when the principal stresses coincide with the propagation directions. When body waves propagate inclined to the principal stress directions in an elastic medium, as discussed previously, we can identify the existence of shear strains in the plane along and/or perpendicular to the wave propagation direction. Hence, Example 2 and 3 will illustrate two scenarios depicting the impact of shear strains on wave velocities. Example 2 examines situations where body waves travel perpendicular to the shear deformation plane, commonly known as acoustoelastic birefringence. In this example, obtaining analytical solutions is feasible. Example 3 involves body waves traveling on the shear deformation plane, posing challenges in deriving analytical forms of acoustoelasticity. Consequently, numerical calculations will be employed to evaluate the effects of shear strain on body wave acoustoelasticity in this scenario. Example 2 and 3 are given in Section 4.2 and Section 4.3, respectively.

### 4.1 Example 1: waves propagating parallel or perpendicular to the principal deformations – existing acoustoelastic expressions in the literature

Assuming that three principal stresses are applied along $x$-, $y$- and $z$-axes, respectively [1, 3-5, 15]. The principal strains that are induced by these principal stresses are denoted as $e_1$, $e_2$ and $e_3$ along $x$-, $y$- and $z$-axes, respectively. Assuming body waves propagate along the $x$-axis in this example. Therefore, Eq. (6) can be expressed as:

$$\begin{bmatrix} B_{1111} & 0 & 0 \\ 0 & B_{2121} & 0 \\ 0 & 0 & B_{3131} \end{bmatrix} \begin{bmatrix} U_1 \\ U_2 \\ U_3 \end{bmatrix} = \rho^0 v^2 \begin{bmatrix} U_1 \\ U_2 \\ U_3 \end{bmatrix}, \qquad (8)$$

where $B_{1111}$, $B_{2121}$ and $B_{3131}$ are shown in Eq. (7).

The diagonal elements are zero since only the principal stresses are involved. As the $B$-matrix in Eq. (8) is diagonal, its three eigenvectors are (1;0;0), (0;1;0), and (0;0;1). Correspondingly, the three eigenvalues are $B_{1111}$, $B_{2121}$ and $B_{3131}$, and three velocities are $(B_{1111}/\rho^0)^{1/2}$, $(B_{2121}/\rho^0)^{1/2}$ and $(B_{3131}/\rho^0)^{1/2}$. Considering that the propagation direction is (1;0;0), the eigenvalue $(B_{1111}/\rho^0)^{1/2}$ corresponds to the velocity of the longitudinal wave, while the $(B_{2121}/\rho^0)^{1/2}$ and $(B_{3131}/\rho^0)^{1/2}$ correspond to the velocities of transverse waves with polarization direction being $y$- and $z$-axes.
In the absence of externally induced strain in the isotropic medium, the expression of wave velocities is reduced to $[(\lambda+2\mu)/\rho^0]^{1/2}$ and $(\mu/\rho^0)^{1/2}$, representing the classic solution of longitudinal and transverse wave velocities, respectively.

## 4.2 Example 2: waves propagating perpendicular to the shear deformation plane – acoustoelastic birefringence

The second example addresses a scenario where the body wave travels perpendicular to the shear deformation plane in an isotropic material. Suppose that the directions of three principal stresses are aligned with the $x$-, $y$- and $z$-axes, respectively, as depicted in Figure 1(a). The corresponding principal strains are denoted as $e_1$, $e_2$ and $e_3$. The three principal stresses can either be tensile or compressive. The wave is propagating along the $x$-axis. The initial governing equation under principal strains is:

$$\begin{bmatrix} B^0_{1111} & 0 & 0 \\ 0 & B^0_{2121} & 0 \\ 0 & 0 & B^0_{3131} \end{bmatrix} \begin{bmatrix} U_1 \\ U_2 \\ U_3 \end{bmatrix} = \rho^0 v^2 \begin{bmatrix} U_1 \\ U_2 \\ U_3 \end{bmatrix}, \tag{9}$$

where $v$ represents the velocity in this stage under principal strains, and the elements in $B^0$-matrix are:

$$B^0_{1111} = \lambda + 2\mu + (5\lambda + 10\mu + 2l + 4m)e_1 + (\lambda + 2l)(e_2 + e_3), \tag{10a}$$

$$B^0_{2121} = \mu + (\lambda + 4\mu + m)e_1 + (\lambda + 2\mu + m)e_2 + \left(\lambda + m - \frac{n}{2}\right)e_3, \tag{10b}$$

$$B^0_{3131} = \mu + (\lambda + 4\mu + m)e_1 + (\lambda + 2\mu + m)e_3 + \left(\lambda + m - \frac{n}{2}\right)e_2. \tag{10c}$$

Please note that the strains involved in Eq. (10) are principal strains, which are different from those in Eq. (7). After establishing a new coordinate system by rotating around the $x$-axis with a certain angle $\alpha$, as shown in Figure 1(b), shear stresses in the $y$-$z$ plane of the new coordinate system are applied. The new strain matrix is calculated as follow:

$$\begin{bmatrix} e_{11} & e_{12} & e_{13} \\ e_{21} & e_{22} & e_{23} \\ e_{31} & e_{32} & e_{33} \end{bmatrix} = \begin{bmatrix} 1 & 0 & 0 \\ 0 & \cos(\alpha) & \sin(\alpha) \\ 0 & -\sin(\alpha) & \cos(\alpha) \end{bmatrix} \begin{bmatrix} e_1 & 0 & 0 \\ 0 & e_2 & 0 \\ 0 & 0 & e_3 \end{bmatrix} \begin{bmatrix} 1 & 0 & 0 \\ 0 & \cos(\alpha) & -\sin(\alpha) \\ 0 & \sin(\alpha) & \cos(\alpha) \end{bmatrix}. \tag{11}$$

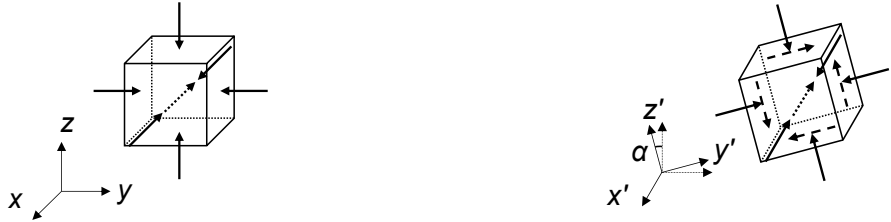

(a) Triaxial principal stress state before coordinate system rotation.

(b) Stress state after transforming to a new coordinate system.

FIGURE 1. Stress states before and after the rotation of the coordinate system in Example 2 (solid arrow: normal stress direction; dashed arrow: shear stress direction; dotted arrow: wave propagation direction).

The new governing equation with the wave propagating along the $x$'-axis (or $x$-axis) after the rotation can be written as:

$$\begin{bmatrix} B_{1111} & 0 & 0 \\ 0 & B_{2121} & B_{3121} \\ 0 & B_{2131} & B_{3131} \end{bmatrix} \begin{bmatrix} U_1 \\ U_2 \\ U_3 \end{bmatrix} = \rho^0 \bar{v}^2 \begin{bmatrix} U_1 \\ U_2 \\ U_3 \end{bmatrix}, \quad (12)$$

where $\bar{v}$ represents the velocity in the new coordinate system, and elements in the *B*-matrix after the coordinate transformation are:

$$B_{1111} = B_{1111}^0, \quad (13a)$$

$$B_{2121} = \mu + (\lambda + 4\mu + m)e_1 + (\lambda + 2\mu + m)\left[\cos^2(\alpha)e_2 + \sin^2(\alpha)e_3\right]$$
$$+ \left(\lambda + m - \frac{n}{2}\right)\left[\cos^2(\alpha)e_3 + \sin^2(\alpha)e_2\right], \quad (13b)$$

$$B_{3131} = \mu + (\lambda + 4\mu + m)e_1 + (\lambda + 2\mu + m)\left[\cos^2(\alpha)e_3 + \sin^2(\alpha)e_2\right]$$
$$+ \left(\lambda + m - \frac{n}{2}\right)\left[\cos^2(\alpha)e_2 + \sin^2(\alpha)e_3\right], \quad (13c)$$

$$B_{2131} = B_{3121} = \left(2\mu + \frac{n}{2}\right)\sin(\alpha)\cos(\alpha)(e_3 - e_2). \quad (13d)$$

The *B*-matrix in Eq. (12) is diagonalizable and can be decomposed as:

$$\begin{bmatrix} B_{1111} & 0 & 0 \\ 0 & B_{2121} & B_{3121} \\ 0 & B_{2131} & B_{3131} \end{bmatrix} = \begin{bmatrix} 1 & 0 & 0 \\ 0 & \cos(\alpha) & \sin(\alpha) \\ 0 & -\sin(\alpha) & \cos(\alpha) \end{bmatrix} \begin{bmatrix} B_{1111}^0 & 0 & 0 \\ 0 & B_{2121}^0 & 0 \\ 0 & 0 & B_{3131}^0 \end{bmatrix} \begin{bmatrix} 1 & 0 & 0 \\ 0 & \cos(\alpha) & -\sin(\alpha) \\ 0 & \sin(\alpha) & \cos(\alpha) \end{bmatrix}. (14)$$

Notably, Eq. (14) is similar to the decomposition of strain matrix in Eq. (11). As shown in Eq. (14), the three eigenvectors are (1;0;0), (0;cos($\alpha$);sin($\alpha$)), and (0;-sin($\alpha$);cos($\alpha$)), which are exactly the same as those for the strain matrix in Eq. (11). The first one aligns with the propagation direction (1;0;0), which represents the longitudinal wave mode. The latter two are orthogonal to the propagation direction, which are relevant to the two transverse wave modes. Given that the eigenvectors specified in Eq. (11) denote the principal strain directions, strikingly, it becomes clear that in this scenario, the polarization directions of body waves will always align with the principal strain directions because of the acoustoelastic effect.

In addition, as shown in Eq. (14), the eigenvalues of the *B*-matrix are solely determined by the $B^0$-matrix, which is calculated based on the principal strain. This implies that in this given configuration, if the magnitudes and directions of the principal strains and the material elastic constants remain constant, the wave velocity, whether it is for the longitudinal or transverse wave, will remain constant regardless of the rotation of the coordinate system or the magnitude of the shear strain. The difference between the second and third eigenvalues of the *B*-matrix can be determined through:

$$B_{2121}^0 - B_{3131}^0 = \rho_0\left(v_{S2}^2 - v_{S3}^2\right) = \left(2\mu + \frac{n}{2}\right)(e_2 - e_3), \quad (15)$$

where $v_{S2}$ and $v_{S3}$ are the second and third eigenvalue-based transverse wave velocities. Equation (15) can be reproduced by Eq. (44) in the article reported by Pao and Gamer [15] under the condition of the isotropic material. This equation reveals that the difference between the squared velocities of two transverse waves is proportional to the difference in principal strain, which is commonly referred to as the acoustoelastic birefringence [42, 52-54]. The primary application scenario for the associated theory in this example is in slightly anisotropic materials, such as rolled metals [53] and timber [55-57], to decouple the velocity changes induced by material textures and applied stresses [54].

From this specific instance, one might notice that the influence of shear strains on the transverse wave velocity cannot be ignored. This observation stems from two aspects. Firstly, as the coordinate system rotates, the normal strain undergoes alterations while the transverse wave velocity remains constant. If the transverse wave velocities are exclusively dependent on the normal strains, their values cannot be constant during the rotation. Hence, this observation indicates the involvement of shear strains in the process. Secondly, in this scenario, the wave polarization always aligns with principal strain directions. If the effect of shear strains is minimal, the moduli $B_{2131}$ and $B_{3121}$ in Eq. (12) can be disregarded, resulting in a diagonalized $B$-matrix. Consequently, the eigenvectors in this matrix, which are (1;0;0), (0;1;0) and (0;0;1), consistently align with the new coordinate system's axes after rotation. However, in reality, the polarization remains unchanged during coordinate system rotation, underscoring the significance of shear strain influence on the polarization direction.

**4.3 Example 3: waves propagating on the shear deformation plane**

Example 2 presents a scenario where the body wave travels perpendicular to the shear deformation plane within an isotropic material. In this section, we will explore the propagation of body waves on the shear deformation plane under a plane stress state, where the out-of-plane principal stress is negligible. Such a stress state is commonly encountered in structural members like concrete girders [58, 59].

In this example, the analytical analysis will be presented first in Section 4.3.1. As obtaining the analytical solution for acoustoelasticity in this case is challenging, a numerical analysis will be conducted in Section 4.3.2. The numerical analysis utilizes the mechanical properties of concrete. In this section, we assume concrete as an isotropic material at a sufficiently large length scale.

*4.3.1 Analytical analysis*

In the initial state, we assume that the two principal stresses are aligned with the $x$- and $y$-axes, respectively, as depicted in Figure 2(a). The corresponding principal strains along the $x$- and $y$-axes are denoted as $e_1$ and $e_2$, respectively. Please note that there exists another principal strain, $e_3$, directed outward from the surface, whose direction is not shown in Figure 2(a). The wave propagates along the $x$-axis in this example, same with the configuration in Example 1 and 2. The initial governing equation is the same as Eq. (9). The three eigenvalues of the $B^0$-matrix are $B^0_1=B^0_{1111}$, $B^0_2=B^0_{2121}$, and $B^0_3=B^0_{3131}$, whose expressions are shown in Eq. (10). It is apparent that the first eigenvalue, $B^0_1$, correlates with the longitudinal wave velocity, whereas the second and third eigenvalues, $B^0_2$ and $B^0_3$, correspond to the transverse wave velocities. In this example, the shear strains are introduced by rotating the original coordinate system in the $x$-$y$ plane to a new coordinate system $x'$-$y'$ with a specific angle $\theta$, as illustrated in Figure 2(b). Since the direction of principal strain $e_3$ is along the axis of rotation, its value remains constant during the rotation. Then, the simplified transformation of the strain matrix can be obtained through:

$$\begin{bmatrix} e_{11} & e_{12} \\ e_{21} & e_{22} \end{bmatrix} = \begin{bmatrix} \cos(\theta) & \sin(\theta) \\ -\sin(\theta) & \cos(\theta) \end{bmatrix} \begin{bmatrix} e_1 & 0 \\ 0 & e_2 \end{bmatrix} \begin{bmatrix} \cos(\theta) & -\sin(\theta) \\ \sin(\theta) & \cos(\theta) \end{bmatrix}, \quad (16)$$

where $\theta$ is the angle between the $x$-axis and the new $x'$-axis after rotation.

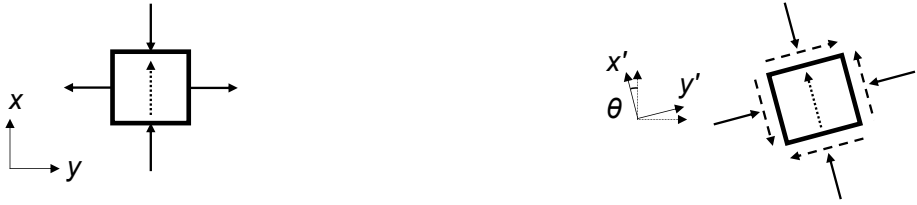

(a) Biaxial principal stress state before coordinate system rotation.  (b) Stress state after transforming to a new coordinate system.

FIGURE 2. Stress states before and after the rotation of the coordinate system in Example 3 (solid arrow: normal stress direction; dashed arrow: shear stress direction; dotted arrow: wave propagation direction).

In the new coordinate system, the wave will propagate along the new $x'$-axis, and the new governing equation is:

$$\begin{bmatrix} B_{1111} & B_{2111} & 0 \\ B_{1121} & B_{2121} & 0 \\ 0 & 0 & B_{3131} \end{bmatrix} \begin{bmatrix} U_1 \\ U_2 \\ U_3 \end{bmatrix} = \rho^0 \bar{v}^2 \begin{bmatrix} U_1 \\ U_2 \\ U_3 \end{bmatrix}, \qquad (17)$$

where the acoustoelastic moduli in the $B$-matrix are shown in Eq. (7). The configuration shown in Eq. (17) is similar to that presented by Destrade and Ogden [60] (please see Section 6.3 of their article).

Since the $B$-matrix in Eq. (17) is symmetric, the analytical solutions for three eigenvalues of this matrix are:

$$\begin{aligned} B_1 &= \frac{B_{1111} + B_{2121} + \sqrt{(B_{1111} - B_{2121})^2 + 4B_{2111}^2}}{2} \\ &= B_{1111} + \frac{\sqrt{1 + \frac{4B_{2111}^2}{(B_{1111} - B_{2121})^2}} - 1}{2}(B_{1111} - B_{2121}), \end{aligned} \qquad (18a)$$

$$\begin{aligned} B_2 &= \frac{B_{1111} + B_{2121} - \sqrt{(B_{1111} - B_{2121})^2 + 4B_{2111}^2}}{2} \\ &= B_{2121} - \frac{\sqrt{1 + \frac{4B_{2111}^2}{(B_{1111} - B_{2121})^2}} - 1}{2}(B_{1111} - B_{2121}), \end{aligned} \qquad (18b)$$

$$B_3 = B_{3131}. \qquad (18c)$$

The first and second eigenvalues comprise the diagonal elements in the $B$-matrix, such as $B_{1111}$ and $B_{2121}$, as well as a term related to shear deformation, $B_{2111}$. This indicates that the eigenvalues are influenced by both normal and shear deformations in the new coordinate system, whereas the diagonal elements are only related to normal deformations in the new coordinate system. Given that the third eigenvalue is equal to the third diagonal element in the $B$-matrix, it will not be further elaborated on in the subsequent discussion. Considering that the direction of wave propagation is along (1;0;0), it is probable that that $B_1$ and $B_{1111}$ correspond to longitudinal waves, whereas $B_2$ and $B_{2121}$ are associated with transverse waves. This hypothesis will be substantiated in the Section 4.3.2.

To better understand the impact of shear deformation-related terms on changes in eigenvalues during coordinate transformations, the difference between the same eigenvalue before and after the coordinate transformation is calculated using:

$$B_1 - B_1^0 = B_{1111} - B_{1111}^0 + \frac{\sqrt{1 + \frac{4B_{2111}^2}{(B_{1111} - B_{2121})^2}} - 1}{2}(B_{1111} - B_{2121}) \;, \tag{19a}$$

$$B_2 - B_2^0 = B_{2121} - B_{2121}^0 - \frac{\sqrt{1 + \frac{4B_{2111}^2}{(B_{1111} - B_{2121})^2}} - 1}{2}(B_{1111} - B_{2121}) \;. \tag{19b}$$

As depicted in Eq. (19), the change in eigenvalue can be separated into two components: the change in the diagonal element $B_{1111}$ or $B_{2121}$ due to the alteration in the normal strains, and the change in the term related to shear deformations.

Exact velocities can be obtained from the eigenvalues of the $B$-matrix using $(B_1/\rho^0)^{1/2}$ and $(B_2/\rho^0)^{1/2}$. These velocities are influenced by both normal and shear deformations. Wave velocities can also be estimated from the diagonal elements in the $B$-matrix using $(B_{1111}/\rho^0)^{1/2}$ and $(B_{2121}/\rho^0)^{1/2}$. These approximate velocities depend on normal deformations. Utilizing the exact and approximate velocities, we can then calculate four velocity changes using the following equations based on the diagonal elements and the eigenvalues to assess potential discrepancies between exact and approximate velocities:

$$\frac{dv}{v}_{P,app} = \frac{v_{P,app} - v_{P,0}}{v_{P,0}} = \frac{\sqrt{\frac{B_{1111}}{\rho^0}} - \sqrt{\frac{\lambda + 2\mu}{\rho^0}}}{\sqrt{\frac{\lambda + 2\mu}{\rho^0}}} = \frac{\sqrt{B_{1111}} - \sqrt{\lambda + 2\mu}}{\sqrt{\lambda + 2\mu}} \;, \tag{20a}$$

$$\frac{dv}{v}_{P,exa} = \frac{v_{P,exa} - v_{P,0}}{v_{P,0}} = \frac{\sqrt{\frac{B_1}{\rho^0}} - \sqrt{\frac{\lambda + 2\mu}{\rho^0}}}{\sqrt{\frac{\lambda + 2\mu}{\rho^0}}} = \frac{\sqrt{B_1} - \sqrt{\lambda + 2\mu}}{\sqrt{\lambda + 2\mu}} \;, \tag{20b}$$

$$\frac{dv}{v}_{S,app} = \frac{v_{S,app} - v_{S,0}}{v_{S,0}} = \frac{\sqrt{\frac{B_{2121}}{\rho^0}} - \sqrt{\frac{\mu}{\rho^0}}}{\sqrt{\frac{\mu}{\rho^0}}} = \frac{\sqrt{B_{2121}} - \sqrt{\mu}}{\sqrt{\mu}} \;, \tag{20c}$$

$$\frac{dv}{v}_{S,exa} = \frac{v_{S,exa} - v_{S,0}}{v_{S,0}} = \frac{\sqrt{\frac{B_2}{\rho^0}} - \sqrt{\frac{\mu}{\rho^0}}}{\sqrt{\frac{\mu}{\rho^0}}} = \frac{\sqrt{B_2} - \sqrt{\mu}}{\sqrt{\mu}} \;, \tag{20d}$$

where d$v$/$v_{P,app}$, d$v$/$v_{P,exa}$, d$v$/$v_{S,app}$ and d$v$/$v_{S,exa}$ represent the approximate velocity changes of longitudinal wave using the first diagonal element $B_{1111}$, the exact velocity changes of longitudinal wave calculated using the first eigenvalue $B_1$, the approximate velocity changes of transverse wave using the second diagonal element $B_{2121}$, and the exact velocity changes of transverse wave calculated using the second eigenvalue $B_2$, respectively.

Unlike Example 1 and 2, it is not feasible to draw any conclusions regarding the impact of shear strains on the wave mode and its corresponding velocity solely through analytical analysis in Example 3. Hence, a numerical analysis is required.

### 4.3.2 Numerical analysis

In the numerical analysis, we will demonstrate the effect of shear strain to the wave velocity with a realistic stress state of a typical concrete structural member. To induce large shear stress during the rotation of the coordinate system, the compressive stress $\sigma_1$ is applied along the $x$-axis with a magnitude of 40% of the compressive strength as one of the principal stresses, as shown in Figure 2(a). It should be noted that the concrete is still considered to be in the elastic stage under this condition [61]. Considering the low concrete tensile strength, in the $y$-axis a tensile stress of 3 MPa is applied. The influence of wave propagation angle $\theta$ to the wave velocity is studied by rotating the angle $\theta$ from 0° to 90° with a step of 5°, while maintaining the principal stresses unchanged. The shear stress arises immediately as the coordinate rotation commences and reaches its maximum at an angle of 45°. The Lamé parameters and Murnaghan constants of concrete used in the numerical analysis are from the paper reported by Nogueira and Rens [62].

The static strain is determined by dividing the static stress by the elastic modulus or shear modulus. Please be aware here that the strain we are using is the tensorial strain, which implies that $e_{ij}=\sigma_{ij}/E$ ($i=j$) and $e_{ij}=1/2\times\sigma_{ij}/G$ ($i\neq j$). This paper exclusively presents the results based on the mechanical properties of Specimen 1 in the article reported by Nogueira and Rens [62]. The compressive strength of Specimen 1 is 33.1 MPa, and the magnitude of applied compressive stress is 40% × 33.1 MPa = 13.24 MPa.

To investigate the impact of shear strains on the first eigenvalue $B_1$, we calculate the value of the first ($B_{1111}-B^0_{1111}$) and the second (the shear deformation-related term) terms in Eq. (19a) relative to the principal stress state before the coordinate rotation, as shown in Figure 3. The horizontal axis represents the angle between the wave propagation direction and the $x$-axis of the coordinate system. As the coordinate system rotates, the term $B_{1111}-B^0_{1111}$ (depicted by the solid black line in Figure 3) are much more significant than those in the shear deformation-related term (represented by the dash-dotted black line in Figure 3). Consequently, the variation in the first eigenvalue, $B_1$, is primarily influenced by the changes in $B_{1111}$. This is evident from the nearly identical absolute values of $B_{1111}$ (shown by the dotted grey line in Figure 3) and $B_1$ (displayed by the dashed grey line in Figure 3). Since $B_{1111}$ is solely related to the normal strains, it can be concluded that the variation in shear strains has minimal impact on $B_1$.

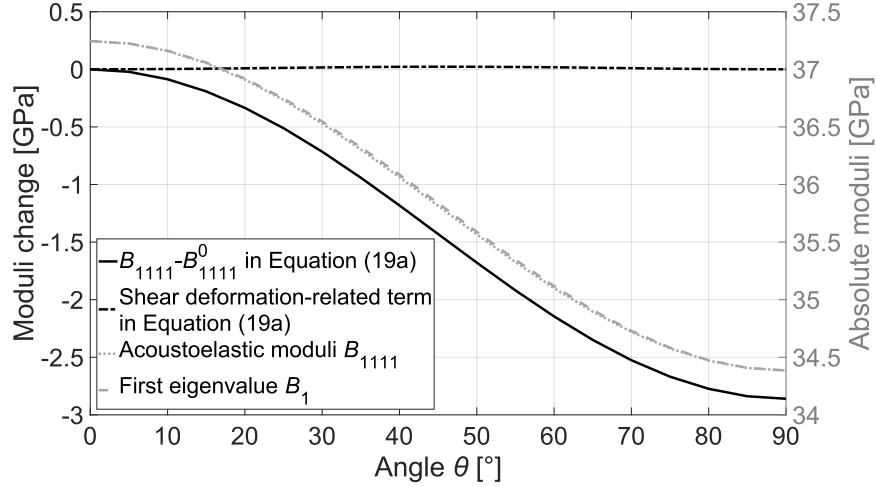

FIGURE 3. Changes in $B_{1111}$ and shear deformation-related term calculated based on Eq. (19a), and the comparison between the moduli $B_{1111}$ and $B_1$ with respect to the rotation angle $\theta$.

During the coordinate transformation, not only the first eigenvalue is changing, its corresponding eigenvector changes as well. Figure 4 displays the first and second scalars in the first eigenvector, while the third scalar, which remains constant at 0, is not included. The first scalar gradually decreases as the shear strain emerges, yet it remains significantly larger than the second scalar and close to 1. Considering that the wave propagates in the direction of (1;0;0), the first eigenvector is not perfectly parallel with the propagation direction. This phenomenon was first reported by Norris [63] and Pao & Gamer [15] in their articles, and this type of wave is referred to as a *quasi-longitudinal wave*, and subsequent researchers have adopted this terminology [45, 64]. Therefore, the wave type linked to the first eigenvalue is that of longitudinal waves.

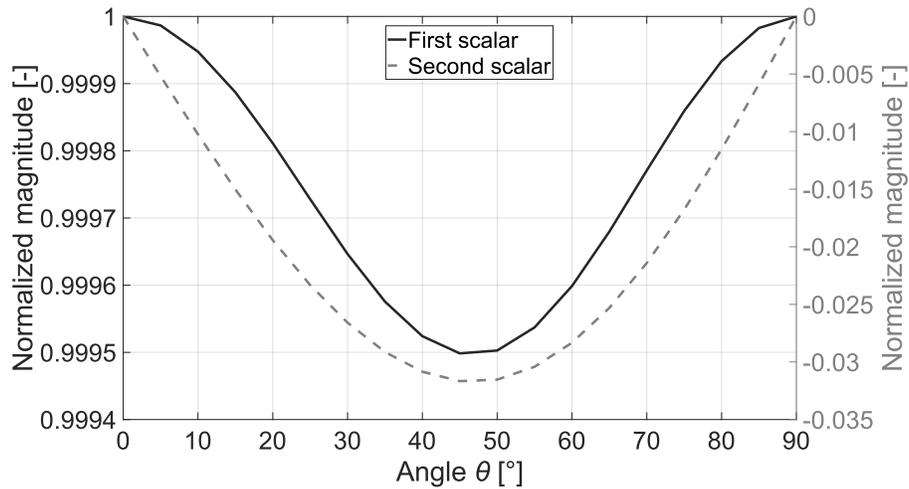

FIGURE 4. The first and second scalars in the first eigenvector of the *B*-matrix in Eq. (17) with respect to the rotation angle $\theta$.

Figure 5 presents the numerical results of the first ($B_{2121}-B^0_{2121}$) and the second (the shear deformation-related term) terms in Eq. (19b) relative to the principal stress state before the coordinate rotation. Unlike the previous findings in Figure 3, the magnitude of the shear deformation-related term (represented by the dash-dotted black line in Figure 5) is comparable to that of $B_{2121}-B^0_{2121}$ (depicted by the solid black line in Figure 5). Consequently, there is a noticeable discrepancy between $B_{2121}$ (the dotted grey line in Figure 5) and $B_2$ (the dashed grey line in Figure 5). Therefore, the second eigenvalue $B_2$ cannot be estimated through $B_{2121}$. Considering that $B_{2121}$ is solely related to the normal strains, one can conclude that shear strains have significant impact on the magnitude of $B_2$. Moving on to Figure 6, we examine the first and second scalars in the second eigenvector of the *B*-matrix during the coordinate transformation. Similar to the observations in Figure 6, it is evident that the wave does not conform to a pure transverse wave mode. This particular type of transverse wave is commonly referred to as a *quasi-transverse wave* [15]. Therefore, the wave type linked to the second eigenvalue is that of transverse waves.

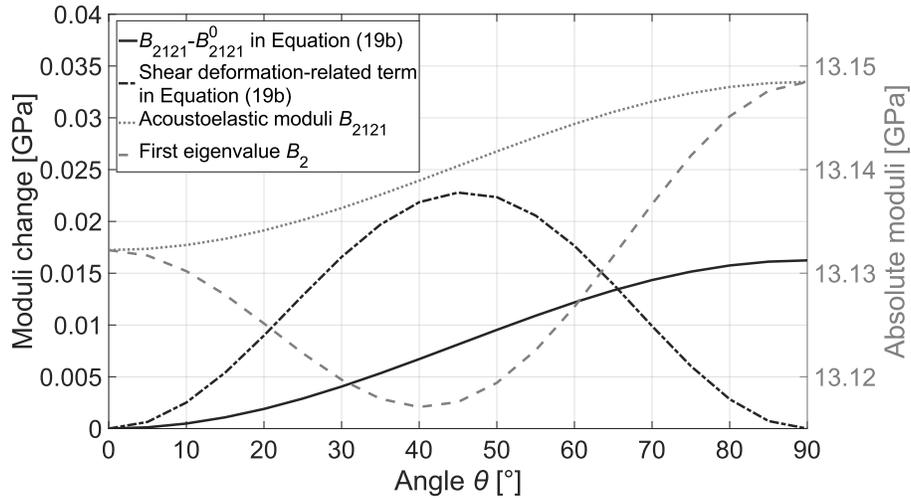

FIGURE 5. Changes in $B_{2121}$ and the shear deformation-related term calculated based on Eq. (19b), and the comparison between the moduli $B_{2121}$ and $B_2$ with respect to the rotation angle $\theta$.

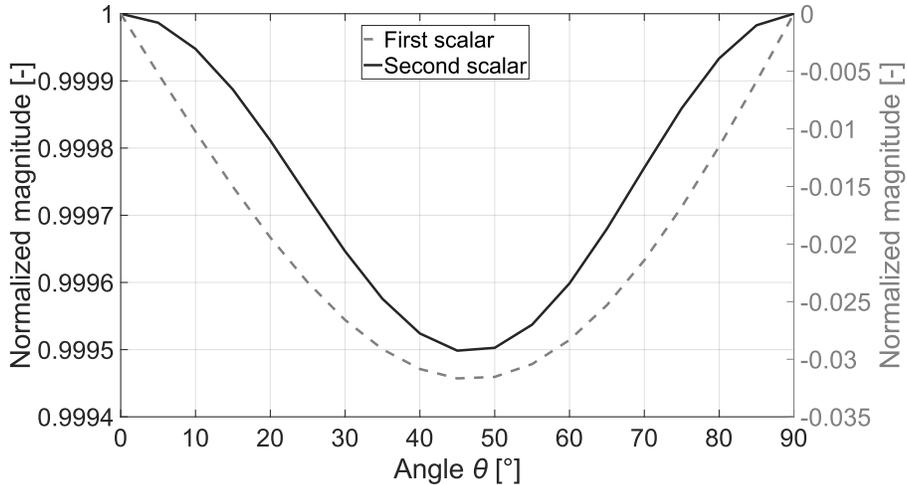

FIGURE 6. The first and second scalars in the second eigenvector of the *B*-matrix in Eq. (17) with respect to the rotation angle $\theta$.

The aforementioned results are derived from the elements and eigenvalues of the *B*-matrix. However, our primary interest lies in the wave velocity obtained from acoustoelasticity. Figure 7 presents the velocity change calculated using Eq. (20). Consistent with the observations in Figure 3 and Figure 5, the velocity change of the quasi-longitudinal wave can be accurately estimated using the diagonal element $B_{1111}$. Additionally, it can be seen from Figure 7 that the minimum exact transverse wave velocity does not occur along the principal directions. A similar phenomenon was reported by Gower et al. [48], who also observed that the highest or lowest wave velocity does not always appear in the principal directions in metals and polymers.

The maximum error occurs when the shear stress reaches its maximum, with an error magnitude of approximately 0.32‰. Considering the magnitudes of applied principal stresses, -13.24 MPa and 3 MPa, this difference is very limited. Therefore, it can be inferred that the longitudinal wave velocity change can be accurately estimated using the first diagonal element $B_{1111}$ in the $B$-matrix, where only normal strains are considered. This inference further suggests that shear strains have a limited effect on longitudinal wave velocities. In contrast, the transverse wave velocity cannot be estimated using the diagonal element $B_{2121}$, confirming our earlier observation from Figure 5, which further suggests that shear strains have a significant effect on transverse wave velocities.

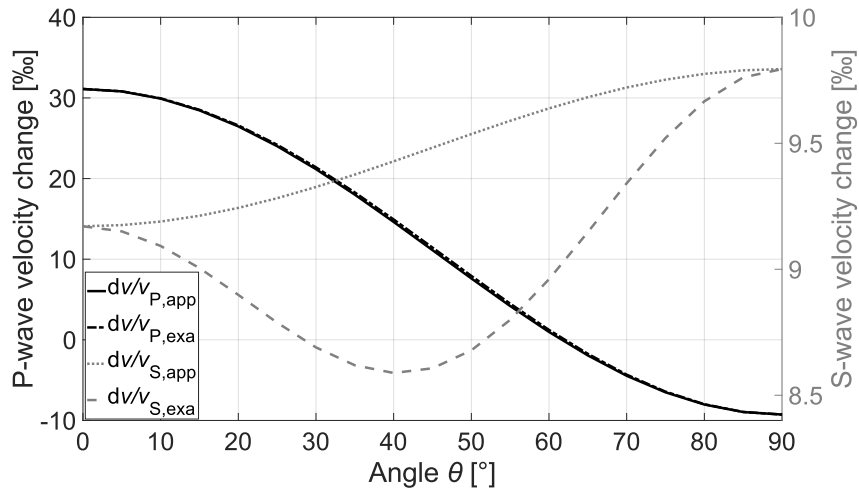

FIGURE 7. Velocity changes of the quasi-longitudinal wave based on the first diagonal element $B_{1111}$, the quasi-longitudinal wave based on the first eigenvalue $B_1$, the quasi-transverse wave based on the second diagonal element $B_{2121}$ and quasi-transverse wave based on the second eigenvalue $B_2$ using Eq. (20) with respect to the rotation angle $\theta$.

## 5. Discussion

We proved in this article that in the stress/strain configuration shown in Example 3, the shear deformation does not have a substantial effect on the longitudinal wave velocity but can significantly change the transverse wave velocity in the context of concrete. In other words, the longitudinal wave velocity depends mainly on the normal stress/strain, while the transverse wave velocity depends on both normal and shear stress/strain. The conclusion regarding longitudinal waves suggests that the acoustoelastic effect of these waves is predominantly associated with elongations or contractions per unit length along the propagation direction, making its function comparable to that of strain gauges. For strain gauges, it is possible to determine the magnitudes and directions of the principal strains in a plane stress state by aligning three strain gauges at different angles. This principle can also be applied to the longitudinal wave-based acoustoelastic effect to determine the magnitudes and directions of the principal strains in a plane stress state. This topic will be further explored in our follow-up studies.

Another intriguing aspect to consider is the influence of concrete heterogeneity—resulting from multiple phases including aggregates, air bubbles, and mortar—on acoustoelasticity. The conclusions presented in this paper are based on the assumption that concrete is an isotropic material. While this treatment of concrete as an isotropic material resembles the concept of *effective medium*, wherein a disordered medium is perceived as a homogeneous effective medium by the probing wave [65], its validity is restricted to coherent waves. Coherent waves are those whose energy predominantly propagates in the initial direction and undergoes limited scattering events [66]. Consequently, the current theoretical model becomes inapplicable for multiply scattered waves. Nevertheless, numerous studies have demonstrated that multiply scattered waves are very sensitive to weak changes in the medium [67], indicating their potential for detecting stress changes in concrete [68, 69]. Therefore, incorporating multiply scattered waves into the current acoustoelastic framework holds promise for enhancing the applicability of this theory in monitoring stress in concrete structures. Bonilla and Keller [70] offer a theoretical examination of how scattering affects acoustoelastic theory, potentially aiding in resolving this issue. We intend to publish a following up study on this topic, exploring the influence of heterogeneity on acoustoelasticity in concrete.

We noted some inconsistent results in the existing literature on acoustoelastic theory, which could lead to inaccurate descriptions of the acoustoelastic effect and misinterpretations of experimental results. For example, Eq. (4a) in the work of Egle and Bray [71] cannot be derived from expressions reported by Hughes and Kelly [3]. Researchers are advised to exercise caution and examine the derivations in cited works to avoid potential inaccuracies when adopting expressions or Murnaghan constants from the literature.

## 6. Conclusion

This paper investigates variations in body wave velocities within a compressible isotropic material subjected to shear and normal deformations. The wave mode and corresponding velocity are determined by solving an eigenvalue problem, where shear strains are embedded in the non-diagonal elements of the acoustoelastic moduli matrix. First, we reproduce the acoustoelasticity expressions for body waves propagating either parallel or perpendicular to the principal axes of deformation, as demonstrated in Example 1. To explore the impact of shear deformation-related terms on wave velocities, we then examine two additional specific examples.

In Example 2, where waves propagate perpendicular to the shear deformation plane, we reaffirm the notion of acoustoelastic birefringence, where wave velocities are solely determined by the magnitude of principal strains. This observation highlights the significance of shear strain in influencing the velocities and polarizations of transverse waves, which cannot be disregarded.

Moving on to Example 3, where waves propagate on the shear deformation plane, we find that the longitudinal wave velocity is primarily influenced by the magnitude of normal strains, with negligible impacts from the shear strains. Conversely, the transverse wave velocity experiences a significant influence from the shear strains, indicating a strong dependency on both normal and shear strains in concrete.

Based on these examples, we draw a general conclusion regarding the influence of shear strains on body wave velocities in concrete: shear strains exert a limited effect on longitudinal wave velocities but significantly alters transverse wave velocities in the context of concrete when body waves propagate on the shear deformation plane.

**Appendix A. Derivations of acoustoelastic moduli in the initial frame**

The derivation details of elements in the *B*-matrix are shown in this appendix. The elements in the *B*-matrix are given by:

$$B_{ijkl} = C_{jlmn}e_{mn}\delta_{ik} + C_{ijkl} + C_{ijklmn}e_{mn}$$
$$+ C_{mjkl}\frac{\partial u_i^{\text{initial}}}{\partial a_m} + C_{imkl}\frac{\partial u_j^{\text{initial}}}{\partial a_m} + C_{ijml}\frac{\partial u_k^{\text{initial}}}{\partial a_m} + C_{ijkm}\frac{\partial u_l^{\text{initial}}}{\partial a_m} \quad . \tag{2}$$

In the following derivation, the linear part of the strain will be used to simplify the expressions:

$$e_{\alpha\beta} = \frac{1}{2}\left(\frac{\partial u_\beta^{\text{initial}}}{\partial a_\alpha} + \frac{\partial u_\alpha^{\text{initial}}}{\partial a_\beta}\right) \quad . \tag{A1}$$

The shear strains in this context exhibit a characteristic shared by all strain definitions and often lead to confusion and errors. Please be aware that the tensorial shear strain in Eq. (A1) is equal to half of the engineering shear strain. The second- and third-order elastic coefficients will be represented using Voigt notation. Here are the derivation details:

$$\begin{aligned}
B_{1111} &= C_{11mn}e_{mn}\delta_{11} + C_{1111} + C_{1111mn}e_{mn} \\
&\quad + C_{m111}\frac{\partial u_1^{\text{initial}}}{\partial a_m} + C_{1m11}\frac{\partial u_1^{\text{initial}}}{\partial a_m} + C_{11m1}\frac{\partial u_1^{\text{initial}}}{\partial a_m} + C_{111m}\frac{\partial u_1^{\text{initial}}}{\partial a_m} \\
&= C_{1111} + C_{1111}e_{11} + C_{1122}e_{22} + C_{1133}e_{33} + C_{111111}e_{11} + C_{111122}e_{22} + C_{111133}e_{33} \\
&\quad + C_{1111}\frac{\partial u_1^{\text{initial}}}{\partial a_1} + C_{1111}\frac{\partial u_1^{\text{initial}}}{\partial a_1} + C_{1111}\frac{\partial u_1^{\text{initial}}}{\partial a_1} + C_{1111}\frac{\partial u_1^{\text{initial}}}{\partial a_1} \\
&= C_{11} + (5C_{11} + C_{111})e_{11} + (C_{12} + C_{112})e_{22} + (C_{13} + C_{113})e_{33} \quad ,
\end{aligned} \tag{A2}$$

$$\begin{aligned}
B_{2121} &= C_{11mn}e_{mn}\delta_{22} + C_{2121} + C_{2121mn}e_{mn} \\
&\quad + C_{m121}\frac{\partial u_2^{\text{initial}}}{\partial a_m} + C_{2m21}\frac{\partial u_1^{\text{initial}}}{\partial a_m} + C_{21m1}\frac{\partial u_2^{\text{initial}}}{\partial a_m} + C_{212m}\frac{\partial u_1^{\text{initial}}}{\partial a_m} \\
&= C_{2121} + C_{1111}e_{11} + C_{1122}e_{22} + C_{1133}e_{33} + C_{212111}e_{11} + C_{212122}e_{22} + C_{212133}e_{33} \\
&\quad + C_{2121}\frac{\partial u_2^{\text{initial}}}{\partial a_2} + C_{2121}\frac{\partial u_1^{\text{initial}}}{\partial a_1} + C_{2121}\frac{\partial u_2^{\text{initial}}}{\partial a_2} + C_{2121}\frac{\partial u_1^{\text{initial}}}{\partial a_1} \\
&= C_{66} + (C_{11} + 2C_{66} + C_{661})e_{11} + (C_{12} + 2C_{66} + C_{662})e_{22} + (C_{13} + C_{663})e_{33} \quad ,
\end{aligned} \tag{A3}$$

$$B_{3131} = C_{11mn}e_{mn}\delta_{33} + C_{3131} + C_{3131mn}e_{mn}$$
$$+ C_{m131}\frac{\partial u_3^{\text{initial}}}{\partial a_m} + C_{3m31}\frac{\partial u_1^{\text{initial}}}{\partial a_m} + C_{31m1}\frac{\partial u_3^{\text{initial}}}{\partial a_m} + C_{313m}\frac{\partial u_1^{\text{initial}}}{\partial a_m}$$
$$= C_{3131} + C_{1111}e_{11} + C_{1122}e_{22} + C_{1133}e_{33} + C_{313111}e_{11} + C_{313122}e_{22} + C_{313133}e_{33} \quad \text{(A4)}$$
$$+ C_{3131}\frac{\partial u_3^{\text{initial}}}{\partial a_3} + C_{3131}\frac{\partial u_1^{\text{initial}}}{\partial a_1} + C_{3131}\frac{\partial u_3^{\text{initial}}}{\partial a_3} + C_{3131}\frac{\partial u_1^{\text{initial}}}{\partial a_1}$$
$$= C_{55} + (C_{11} + 2C_{55} + C_{551})e_{11} + (C_{12} + C_{552})e_{22} + (C_{13} + 2C_{55} + C_{553})e_{33} \; ,$$

$$B_{1121} = C_{11mn}e_{mn}\delta_{12} + C_{1121} + C_{1121mn}e_{mn}$$
$$+ C_{m121}\frac{\partial u_1^{\text{initial}}}{\partial a_m} + C_{1m21}\frac{\partial u_1^{\text{initial}}}{\partial a_m} + C_{11m1}\frac{\partial u_2^{\text{initial}}}{\partial a_m} + C_{112m}\frac{\partial u_1^{\text{initial}}}{\partial a_m}$$
$$= C_{112112}e_{12} + C_{112121}e_{21} + C_{2121}\frac{\partial u_1^{\text{initial}}}{\partial a_2} + C_{1221}\frac{\partial u_1^{\text{initial}}}{\partial a_2} + C_{1111}\frac{\partial u_2^{\text{initial}}}{\partial a_1} + C_{1122}\frac{\partial u_1^{\text{initial}}}{\partial a_2} \quad \text{(A5)}$$
$$= (2C_{166} + 2C_{11})e_{12} = B_{2111} \; ,$$

$$B_{1131} = C_{11mn}e_{mn}\delta_{13} + C_{1131} + C_{1131mn}e_{mn}$$
$$+ C_{m131}\frac{\partial u_1^{\text{initial}}}{\partial a_m} + C_{1m31}\frac{\partial u_1^{\text{initial}}}{\partial a_m} + C_{11m1}\frac{\partial u_3^{\text{initial}}}{\partial a_m} + C_{113m}\frac{\partial u_1^{\text{initial}}}{\partial a_m}$$
$$= C_{113113}e_{13} + C_{113131}e_{31} + C_{3131}\frac{\partial u_1^{\text{initial}}}{\partial a_3} + C_{1331}\frac{\partial u_1^{\text{initial}}}{\partial a_3} + C_{1111}\frac{\partial u_3^{\text{initial}}}{\partial a_1} + C_{1133}\frac{\partial u_1^{\text{initial}}}{\partial a_3} \quad \text{(A6)}$$
$$= (2C_{155} + 2C_{11})e_{13} = B_{3111} \; ,$$

$$B_{2131} = C_{11mn}e_{mn}\delta_{23} + C_{2131} + C_{2131mn}e_{mn}$$
$$+ C_{m131}\frac{\partial u_2^{\text{initial}}}{\partial a_m} + C_{2m31}\frac{\partial u_1^{\text{initial}}}{\partial a_m} + C_{21m1}\frac{\partial u_3^{\text{initial}}}{\partial a_m} + C_{213m}\frac{\partial u_1^{\text{initial}}}{\partial a_m}$$
$$= C_{213123}e_{23} + C_{213132}e_{32} + C_{3131}\frac{\partial u_2^{\text{initial}}}{\partial a_3} + C_{2121}\frac{\partial u_3^{\text{initial}}}{\partial a_2} \quad \text{(A7)}$$
$$= (2C_{654} + 2C_{44})e_{23} = B_{3121} \; .$$